\documentclass[preprint,fleqn,3p]{elsarticle}
\usepackage{graphicx}
\usepackage{pifont}
\usepackage{natbib}
\usepackage{geometry}

\usepackage{makeidx}  % allows for indexgeneration
\usepackage{times}
\usepackage{amsfonts}
\usepackage{amssymb}
\usepackage{amsmath}
\usepackage{xspace}
\usepackage{subfigure}

\usepackage[dvipdfm]{hyperref}

\journal{Computer Physics Communications}

\begin{document}
\begin{frontmatter}
\cortext[cor1]{Corresponding author. \\E-mail address: fukuko.yuasa@kek.jp (F.Yuasa), Tel: +81-29-879-6009, Fax: +81-29-864-4402, Postal address: 1-1 Oho Tsukuba Ibaraki 305-0801 Japan.}

\title{Numerical Computation of Two-loop Box Diagrams with Masses}

\author[KEK]{F. Yuasa\corref{cor1}}
\author[WMU]{E. de Doncker} 
\author[KEK]{N. Hamaguchi} 
\author[KEK]{T. Ishikawa} 
\author[KOG]{K. Kato} 
\author[KEK]{Y. Kurihara} 
\author[KEK]{J. Fujimoto} 
\author[KEK]{Y. Shimizu}

\address[KEK]{High Energy Accelerator Research Organization (KEK), 1-1 Oho Tsukuba, Ibaraki 305-0801, Japan}
\address[WMU]{Western Michigan University Kalamazoo, MI 49008-5371, USA}
\address[KOG]{Kogakuin University, 1-24 Nishi-Shinjuku, Shinjuku, Tokyo 163-8677, Japan}

\begin{abstract}
A new approach is presented to evaluate multi-loop
integrals, which appear in the calculation of cross-sections
in high-energy physics. It relies on a fully numerical method and is
applicable to a wide class of integrals with various mass
configurations. As an example, the computation of two-loop planar
and non-planar box diagrams is shown. The results are confirmed
by comparisons with other techniques, including the reduction method,
and by a consistency check using the dispersion relation.
\end{abstract}


\end{frontmatter}

%-----------------------------------------------------------------------
%%%%%%%%% Section 1  %%%%%%%%%%
\section{Introduction}\label{sec:intro}
In the study of high-energy reactions observed at current and future accelerators, such as LHC and ILC, 
precise theoretical predictions of cross-sections including higher order corrections are required.
This is due to the fact that the lowest order approximation in perturbative calculations of quantum field theory is not
sufficiently accurate to be compared to the experimental data. One has to take into account 
the contributions from higher order terms as well. In order to include these corrections in the Standard model 
or beyond, it is indispensable to handle the evaluation of loop integrals.

At the one-loop level it is known that analytic solutions exist for any type of diagram, and the results 
are expressed in terms of known functions, such as logarithms and Spence functions (see, for example~\cite{FF}). 
Using these analytic results several automatic computation systems~\cite{formcalc,xloop,grace,rocket,helac,madloop,blackhat,golem,golem-samurai} 
have been proposed.  In order to estimate cross sections we need automatic computation systems because we may have 
to deal with a large number of relevant Feynman diagrams for a given process.

However, the extension of the system to include higher order corrections is not an easy task, because
analytic integration is generally impossible for higher loop diagrams, especially for diagrams which depend on more general mass configurations. Analytic results are only known for a limited class of two-loop diagrams.
Therefore we have to rely on numerical evaluations. We need to establish efficient methods that can be incorporated into 
automatic computation systems of cross-sections. For a number of years we have gained experience evaluating one-loop integrals 
numerically, where the results can be compared with known analytic answers. We succeeded in calculating vertex, box and 
pentagon diagrams with arbitrary masses~\cite{dq1,dq2,dq3,dq4,dq5,acat07,acat08,nmms10,jocs11,intech,acat11,qfthep11}. We also computed two-loop 
self-energy and vertex diagrams. Further related work can be found in~\cite{1loop1,supplement,1loop2,1loop3,kreimer,2loopself,2loopv1,2loopv2,tarasov,bauberger,passarino-self,passarino-2lv,nci,anastasiou,nagysoper,ueda-acat08,beckerweinzierl}. 

In our method we start from the Feynman parameter representation of loop integrals. We employ a fully
numerical integration procedure combined with numerical extrapolation. The purpose of this paper is to 
describe the method in detail and to show results for more complicated loop integrals, corresponding to two-loop box diagrams with massive 
particles. For simplicity we deal with scalar loop integrals throughout this paper, ignoring all spin complications 
that are irrelevant to the essential discussion of the numerical approach.

The most general form of the scalar integral for a diagram with $L$ loops and $N$ internal lines is given by
\begin{equation}
{\mathcal I} = % \lim_{\epsilon \rightarrow 0+}
               \prod_{j=1}^{L} \int \frac{d^{n}l_{j}}{i(2\pi)^n}\prod_{r=1}^{N} \frac{1}{D_{r}}
\end{equation}
where $l_j$ is the $j$-th loop momentum in the $n$-dimensional space-time, and
\begin{equation}
D_{r}=q_{r}^2-m_{r}^2+i\epsilon
\end{equation}
is the inverse of the $r$-th Feynman propagator, where $\epsilon$ denotes an infinitesimal quantity,
$m_{r}$ is the mass of the $r$-th particle, and the momentum $q_{r}$ flowing on the $r$-th internal line is given by a sum of 
loop and external momenta. We make use of the Feynman identity,
\begin{equation}
\prod_{r=1}^{N}\frac{1}{D_{r}}=\Gamma(N)\int_{0}^{1}\prod_{r=1}^{N} dx_{r}\,\frac{\delta(1-\sum x_{r})}{(\sum x_{r}D_{r})^N}.
\end{equation}
Carrying out the loop momentum integrations delivers
\begin{equation}
\mathcal I = \left(\frac{1}{4\pi}\right)^{nL/2}\Gamma\left(N-\frac{nL}{2}\right) \times I,
\end{equation}
where
\begin{equation}\label{eq:scalar-integral}
I =  (-1)^N \int_{0}^{1}\prod_{i=1}^{N}dx_{i}\, \delta(1-\sum x_{i})\frac{C^{N-n(L+1)/2}}{(D-i\epsilon C)^{N-nL/2}}. 
\end{equation}
\noindent
The function $D$ is a polynomial in the Feynman parameters $\{x_i\}$. $D$ further involves physical variables such as the external 
momenta and particle masses. The function $C$ is also a polynomial in the $\{x_i\}$. Both functions are determined by 
the topology of the Feynman diagram. Details of their construction are summarized in Appendix A.

In the two-loop box diagrams, $D$ depends on two kinematical variables $s$ and $t$, where $s$ is the square of the total 
energy of the colliding particle system, and $t<0$ is the squared momentum transfer between the initial and the final particles.
For the infrared divergent integrals, we have two prescriptions. 
One is to introduce a small fictitious mass $\lambda$ for the massless particles and the other is the dimensional regularization technique. 
In the former we can set $n=4$ in Eq.(\ref{eq:scalar-integral}) and the procedure is straightforward once the value  $\lambda$ is fixed ~\cite{dq4,acat07}.
%The former has an advantage that there is no extension in our method and the procedure is straightforward once the value  $\lambda$ is fixed ~\cite{dq4,acat07}.  
%Using $\lambda$ we can set $n=4$ in Eq.(\ref{eq:scalar-integral}).
For the latter we put $n=4+2\delta= n(\delta)$ and use {\it a double extrapolation technique} for both $\epsilon$ and $\delta$ (from $n(\delta)$) in Eq.(\ref{eq:scalar-integral})~\cite{intech,acat11}. 
Here we estimate the integral for a fixed value of $\delta$ using the extrapolation with respect to $\epsilon$. 
%Repeating this for a series of $\delta$ values, we can numerically estimate the pole residue of $\frac{1}{\delta}$ and the finite part of the integral.
Repeating this for a series of $\delta$ values, we can estimate the pole residue of $1/\delta$ and the finite part of the integral numerically.
%Repeating this for a series of $\delta$ values, we can numerically estimate the pole residue of $\displaystyle{\frac{1}{\delta}}$ and the finite part of the integral.
%When we take the former, we can set $n=4$ in Eq.(\ref{eq:scalar-integral}) for the ultra-violet finite integrals.

We briefly describe the general properties of the integral $I$ and give some terminology. Depending on the value of 
$s$, the function $D$ in the denominator may vanish in the integration domain. In this case, the infinitesimal parameter 
$\epsilon$ prevents $I$ from diverging. Then $I$ exhibits an imaginary part even if all the physical parameters $s$, 
$t$ and the masses are real. This region of $s$ is called the {\it physical region}, where $s$ exceeds the threshold 
energy, so that the reaction takes place. On the other hand, in the {\it unphysical region},
$s$ is lower than the threshold. This is the region of $s$ where we can put $\epsilon = 0$ and the integral is real 
for real $s$. Thus the integral $I$ can be regarded as an analytic function in the complex $s$-plane with cuts 
along the real $s$-axis, starting at branch points which are determined by physics conditions. However, as we shall 
see below in Section~\ref{sec:tech}, we treat $\epsilon$ not as infinitesimal but as a {\it finite} number in the numerical procedure for calculating
 $\Re(I)$ and $\Im(I)$, $I=\Re(I)+i\Im(I)$, in the physical region. 

This paper is outlined as follows. In Section~\ref{sec:def} we construct the integrands for two-loop box diagrams ($L=2$ 
and $N=7$), and present suitable variable transformations. We explain the details of our techniques in 
Section~\ref{sec:tech}; and the results of the computations are shown in Section~\ref{sec:results}. Section~\ref{sec:quality} is devoted 
to a discussion on how to assess the correctness of the obtained results. Section~\ref{sec:summary} gives conclusions and future directions for this work. 
%-----------------------------------------------------------------------

%%%%%%%%% Section2  %%%%%%%%%%
\section{Two-loop box integrals}\label{sec:def}
\begin{center}
\begin{figure}[htb]
\subfigure[] {
\includegraphics[width=0.45\linewidth]{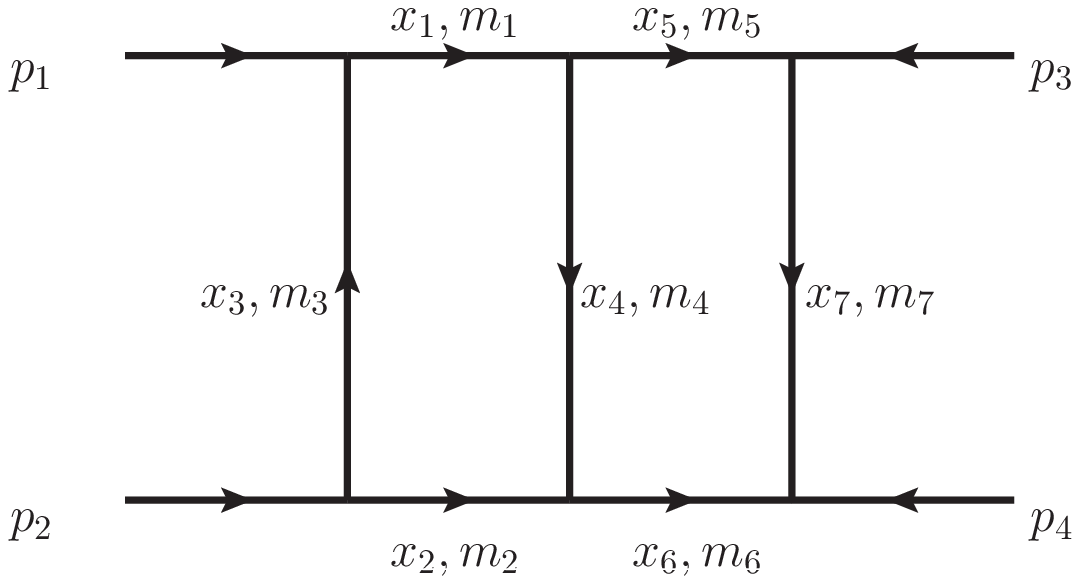}\label{fig:tlb-a}}
\subfigure[] {
\includegraphics[width=0.45\linewidth]{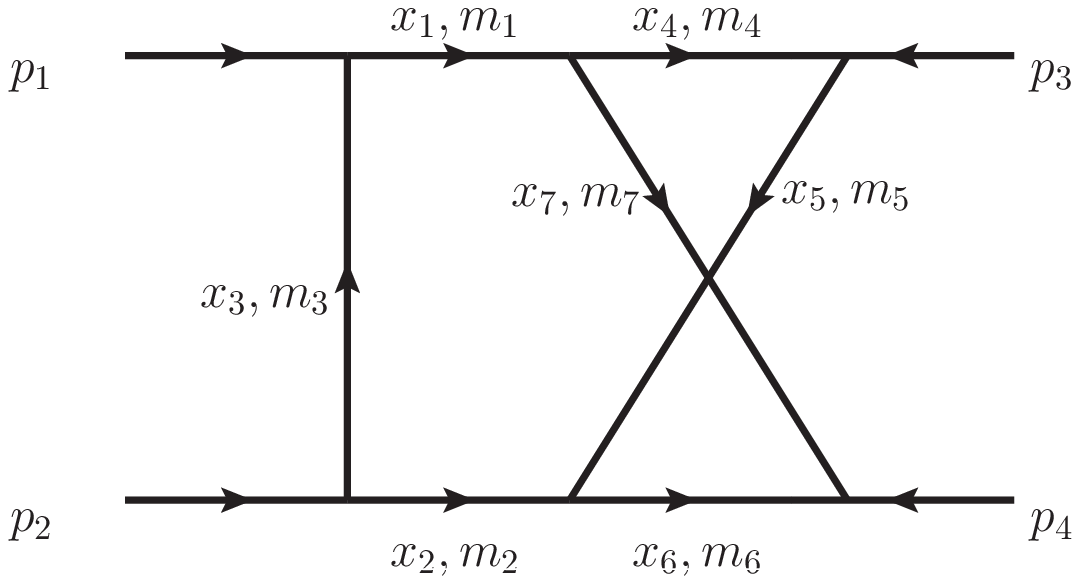}\label{fig:tlb-b}}
\caption{(a) Two-loop planar box diagram ~~ (b) Two-loop non-planar box diagram} 
\end{figure}
\end{center}

The topology of the two-loop box diagram is depicted in Fig.\,1. We call Fig.~\ref{fig:tlb-a} the planar diagram and Fig.~\ref{fig:tlb-b} 
the non-planar diagram, respectively. The loop integral in the Feynman parameters ($x_1, \cdots , x_7$) is of the form 
\begin{equation}\label{eq:2lb}
I = -\int_{0}^{1} dx_{1} ~dx_{2} ~dx_{3} ~dx_{4} ~dx_{5} ~dx_{6} ~dx_{7} 
 ~\delta(1-\sum_{\ell=1}^{7}x_{\ell})\,\frac{C}{(D-i\epsilon C)^{3}}.
\end{equation}
Here, $D$ and $C$ are polynomials of Feynman parameters. Their derivations are given in \ref{app:DandC}.
%%%%%%%%%%%%%%%%%%%%%%%%%%%%%%%%%%%%%%%%%%%%%%%%%%%%%%
%%%%%%%%%%%%%%%%%%%%%%%%%%%%%%%%%%%%%%%%%%%%%%%%%%%%%%
The external momenta $p_1, p_2, p_3$ and $p_4$ are defined to flow inward, satisfying $p_{1}+p_{2}+p_{3}+p_{4}=0$.
The kinematical variables $s$ and $t$ are given by
\begin{eqnarray*}
s&=&(p_{1}+p_{2})^2=(p_{3}+p_{4})^2,\\
t&=&(p_{1}+p_{3})^2=(p_{2}+p_{4})^2.
\end{eqnarray*}
For later notational convenience we introduce a third kinematical variable $u$ by
\begin{equation*}
u=(p_1+p_4)^2=(p_2+p_3)^2.
\end{equation*}
The variables $s,t$ and $u$ are not independent, as
\begin{equation*}
s+t+u=p_1^2+p_2^2+p_3^2+p_4^2.
\end{equation*}

In the following we derive the explicit formulae of the functions ${D}$ and ${C}$. 
We also show examples of variable transformations, which allow eliminating a common factor in the numerator and denominator. Furthermore, the resulting form of the integral will be suited for an application of the reduction formalism given in Section~\ref{subsec:reduction}. 
Followed by a Monte Carlo integration we will use the latter for the purpose of comparing of its numerical results with those by DCM.
%In the multi-loop case, with these we can eliminate a factor common to both numerator and denominator.
%In principle, DCM can work without specific variable transformation.  
%However, sometimes we can eliminate a factor common to both numerator and denominator. 
%This stabilizes the numerical computation, so that the computation time can be reduced.
%Since the transformation only depends only on the graph topology, it can be applied automatically after preparation of enough number of templates.
%In the process of transformations some common factors appearing in both the numerator and denominator of Eq.(\ref{eq:2lb}) can be devided and this stabilizes the numerical integration.
%An adequate variable transformation can reduce the computation time for the integration and it can be helpful to find the symmetries among variables.
 
\subsection{Explicit formulae of the functions $D$ and $C$ for the planar diagram}\label{subsec:tlbl}
The functions ${D}$ and ${C}$ in Eq.\,(\ref{eq:2lb}), corresponding to Fig.~\ref{fig:tlb-a}, are given by 
\begin{eqnarray}\label{eq:D-ladder}
{D}&=& {C} \sum x_\ell m^2_\ell \\ \nonumber
            &-&\{s (x_{1}x_{2}(x_{4} + x_{5} + x_{6} + x_{7})+x_{5}x_{6}(x_{1} + x_{2} + x_{3} + x_{4}) + x_{1}x_{4}x_{6} + x_{2}x_{4}x_{5}) \\ \nonumber
           &+& t x_{3}x_{4}x_{7} \\ \nonumber
           &+& p_{1}^2(x_{3}(x_{1}x_{4} + x_{1}x_{5} + x_{1}x_{6} + x_{1}x_{7} + x_{4}x_{5})) \\ \nonumber
           &+& p_{2}^2(x_{3}(x_{2}x_{4} + x_{2}x_{5} + x_{2}x_{6} + x_{2}x_{7} + x_{4}x_{6})) \\ \nonumber
           &+& p_{3}^2(x_{7}(x_{1}x_{4} + x_{1}x_{5} + x_{2}x_{5} + x_{3}x_{5} + x_{4}x_{5})) \\ \nonumber
           &+& p_{4}^2(x_{7}(x_{1}x_{6} + x_{2}x_{4} + x_{2}x_{6} + x_{3}x_{6} + x_{4}x_{6}))\},
\end{eqnarray}
and
\begin{equation}\label{eq:C-ladder}
{C}=(x_{1} + x_{2} + x_{3} + x_{4})(x_{4} + x_{5} + x_{6} + x_{7})-x_{4}^2.
\end{equation}

Using the transformation
$(x_{1},x_{2},x_{3},x_{4},x_{5},x_{6},x_{7}) \rightarrow (\rho_1, \rho_2, \rho_3, u_{1},u_{2},u_{3},u_{4}),$ defined by 
$x_{1} = \rho_{1} u_{1}$,
$x_{2} = \rho_{1} u_{2}$,
$x_{3} = \rho_{1} (1- u_{1} - u_{2})$,
%$x_{4} = 1 - \rho_{1} - \rho_{2}$,~~
$x_{4} = \rho_{3}$,
$x_{5} = \rho_{2} u_{3}$,
$x_{6} = \rho_{2} u_{4}$ and
$x_{7} = \rho_{2} (1- u_{3}-u_{4})$,
we obtain
\noindent
$$
\int ~dx_{1} \cdot \cdot \cdot ~dx_{7} ~\delta(1-\sum x_{j}) \\
= \int ~d\rho_{1} ~d\rho_{2} ~d\rho_{3} ~\delta(1-\sum \rho_{j})\, \rho_{1}^2 \rho_{2}^2
\int ~du_{1} ~du_{2} ~du_{3} ~du_{4},
$$
with
$$
~~\rho_{1}+\rho_{2}+\rho_{3}=1, \\
~~0\leq u_{1}+u_{2} \leq 1, ~~0\leq u_{3}+u_{4} \leq 1,
$$
and with Jacobian $\rho_{1}^2 \rho_{2}^2$. Changing the variables by
$
\rho_{1} = \rho \xi, 
\rho_{2} = \rho (1 - \xi)$
and
$
\rho_{3} = 1 - \rho
$
gives
%\begin{equation}
$$\int ~d\rho_{1} ~d\rho_{2} ~d\rho_{3} ~\delta(1-\sum \rho_{j})\, \rho_{1}^2 \rho_{2}^2 \cdots =
\int_{0}^{1} ~d\rho \int_{0}^{1} ~d\xi ~\rho^{5} \xi^{2}(1-\xi)^{2} \cdots.
$$
%\end{equation}
After these transformations, $D$ and $C$ contain a common factor $\rho$ and we set ${\mathcal D} = {D}/\rho$ and ${\mathcal C} = {C}/\rho$.
The integral becomes
\begin{equation}\label{eq:integ-ladder}
{I_{planar}} = -\int_{0}^{1} d{\rho} \int_{0}^{1}d{\xi} \int_{0}^{1} du_{1} \int_{0}^{1-u_{1}}du_{2} \int_{0}^{1} du_{3} \int_{0}^{1-u{3}}du_{4} ~\frac{{\mathcal C}}{({\mathcal D}-i\epsilon {{\mathcal C}})^{3}}\,\rho^{3} \xi^{2
}(1-\xi)^{2},
\end{equation}
where ${\mathcal D}$ is a quadratic in ${\bf u} = (u_1,u_2,u_3,u_4)^T$,
\begin{equation}\label{eq:Dprime-ladder}
{\mathcal D} = {\bf u}^T A {\bf u} + {\bf B}^T {\bf u} + c,
\end{equation}
\noindent
and
\begin{equation}\label{eq:Cprime-ladder}
{\mathcal C}
= \rho \xi (1-\xi) + 1 - \rho.
%= -\rho \xi^2 + \rho \xi - \rho + 1.
\end{equation}

%It should be noted that the integral in $u$'s shows the same form as the pentagon diagram in one-loop.
The $4\times4$ matrix $A$~ is symmetric and depends on the internal masses $m_{\ell}~~(1\le \ell\le 7)$,
and on the kinematical variables, $s$ and $t$. In this paper we assume $p_1^2=p_2^2=p_3^2=p_4^2=m^2$ for both diagrams.
When $m_{1}= m_{2}=m_{5}=m_{6}=m$ and $m_{3}=m_{4}=m_{7}=M ~(m \neq M)$, we have 
$$A =
\begin{pmatrix}
\rho\xi^2(1-\rho\xi)A_{1} ~&~ \rho\xi(1-\rho)(1-\xi)A_{2}\\
\rho\xi(1-\rho)(1-\xi)A_{2} ~&~ \rho(1-\xi)^2(1-\rho(1-\xi))A_{1}\\
\end{pmatrix},
$$
\noindent
where the $2\times2$ matrices $A_{1}$ and $A_{2}$ are
$$A_{1} =
\begin{pmatrix}
-m^2~&~ s/2-m^2\\
s/2-m^2~&~ -m^2\\
\end{pmatrix},
 ~~A_{2} =
\begin{pmatrix}
t/2-m^2~&~ s/2+t/2-m^2\\
s/2+t/2-m^2~&~t/2-m^2 \\
\end{pmatrix}.
$$
\noindent
The vector ${\bf B}$ is given by
$${\bf{B}} =
\begin{pmatrix}
-t \rho\xi(1-\rho)(1-\xi) + M^2 \rho\xi{\mathcal C} \\
-t \rho\xi(1-\rho)(1-\xi) + M^2 \rho\xi{\mathcal C} \\
-t \rho\xi(1-\rho)(1-\xi) + M^2 \rho(1-\xi){\mathcal C} \\
-t \rho\xi(1-\rho)(1-\xi) + M^2 \rho(1-\xi){\mathcal C}
\end{pmatrix},
$$
\noindent
and the scalar $c$ is
\begin{equation}
c = t \rho\xi(1-\rho)(1-\xi) - M^2 {\mathcal C}.\nonumber
\end{equation}
The quadratic form will further be used in Section~\ref{subsec:reduction} for a comparison of DCM with a reduction method. 
\subsection{Explicit formulae of the functions $D$ and $C$ for the non-planar diagram}\label{subsec:tlbc}
The functions $D$ and $C$ in Eq.~(\ref{eq:2lb}), corresponding to Fig.~\ref{fig:tlb-b}, are 
%\begin{eqnarray*}
%{\mathcal D}&=& {\mathcal C} \sum x_\ell m^2_\ell \\
%            &-&\{s (- x_{1}x_{5}x_{6} + x_{2}x_{3}x_{4} + x_{2}x_{3}x_{5} + x_{2}x_{3}x_{6} + x_{2}x_{3}x_{7} + x_{2}x_{6}x_{7} + x_{3}x_{4}x_{5}) \\
%            &+& t (x_{1}(x_{4}x_{7} - x_{5}x_{6}))\\
%            &+& p_{1}^2(x_{1}(x_{2}x_{4} + x_{2}x_{5} + x_{2}x_{6} + x_{2}x_{7} + x_{4}x_{5} + x_{5}x_{6}))\\
%            &+& p_{2}^2(x_{1}(x_{3}x_{4} + x_{3}x_{5} + x_{3}x_{6} + x_{3}x_{7} + x_{5}x_{6} + x_{6}x_{7}))\\
%            &+& p_{3}^2(x_{1}x_{5}x_{6} + x_{1}x_{5}x_{7} + x_{2}x_{4}x_{7} + x_{2}x_{5}x_{7} + x_{3}x_{5}x_{6} + x_{3}x_{5}x_{7} + x_{4}x_{5}x_{7} + x_{5}x_{6}x_{7})\\
%            &+& p_{4}^2(x_{1}x_{4}x_{6} + x_{1}x_{5}x_{6} + x_{2}x_{4}x_{6} + x_{2}x_{5}x_{6} + x_{3}x_{4}x_{6} + x_{3}x_{4}x_{7} + x_{4}x_{5}x_{6} + x_{4}x_{6}x_{7}) \} 
%\end{eqnarray*}
\begin{eqnarray}\label{eq:D-cross}
 D&=& C \sum x_\ell m^2_\ell \\ \nonumber
            &-&\{s (x_{1}x_{2}x_{4} + x_{1}x_{2}x_{5} + x_{1}x_{2}x_{6} + x_{1}x_{2}x_{7} + x_{1}x_{5}x_{6} + x_{2}x_{4}x_{7} - x_{3}x_{4}x_{6}) \\ \nonumber
            &+& t (x_{3}(- x_{4}x_{6} + x_{5}x_{7}))\\ \nonumber
            &+& p_{1}^2(x_{3}(x_{1}x_{4} + x_{1}x_{5} + x_{1}x_{6} + x_{1}x_{7} + x_{4}x_{6} + x_{4}x_{7}))\\ \nonumber
            &+& p_{2}^2(x_{3}(x_{2}x_{4} + x_{2}x_{5} + x_{2}x_{6} + x_{2}x_{7} + x_{4}x_{6} + x_{5}x_{6}))\\ \nonumber
            &+& p_{3}^2(x_{1}x_{4}x_{5} + x_{1}x_{5}x_{7} + x_{2}x_{4}x_{5} + x_{2}x_{4}x_{6} + x_{3}x_{4}x_{5} + x_{3}x_{4}x_{6} + x_{4}x_{5}x_{6} + x_{4}x_{5}x_{7})\\ \nonumber
            &+& p_{4}^2(x_{1}x_{4}x_{6} + x_{1}x_{6}x_{7} + x_{2}x_{5}x_{7} + x_{2}x_{6}x_{7} + x_{3}x_{4}x_{6} + x_{3}x_{6}x_{7} + x_{4}x_{6}x_{7} + x_{5}x_{6}x_{7}) \} 
\end{eqnarray}
and
\begin{equation}\label{eq:C-cross}
C=(x_{1} + x_{2} + x_{3} + x_{4} + x_{5})(x_{1} + x_{2} + x_{3} + x_{6} + x_{7})-(x_{1} + x_{2} + x_{3})^2.
\end{equation}

The transformation
$(x_{1},x_{2},x_{3},x_{4},x_{5},x_{6},x_{7}) \rightarrow (\rho_{1},\rho_{2},\rho_{3},u_{1},u_{2},u_{3},u_{4}),$ defined by 
$x_{1} = \rho_{1} u_{1},$
$x_{2} = \rho_{1} u_{2},$
$x_{3} = \rho_{1} (1-u_{1}-u_{2}),$
$x_{4} = \rho_{2} u_{3},$
$x_{5} = \rho_{2} (1-u_{3}),$
$x_{6} = \rho_{3} u_{4}$ and
$x_{7} = \rho_{3} (1-u_{4}),$
yields
\noindent
%\begin{eqnarray*}
$$
\int ~dx_{1} \cdots ~dx_{7} ~\delta(1-\sum x_{j}) \\
= \int ~d\rho_{1} ~d\rho_{2} ~d\rho_{3} ~\delta(1-\sum \rho_{j})\, \rho_{1}^2 \rho_{2} \rho_{3}
\int ~du_{1} ~du_{2} ~du_{3} ~du_{4},
$$
%\end{eqnarray*}
with
$$
~~\rho_{1}+\rho_{2}+\rho_{3}=1, \\
~~0\leq u_{1}+u_{2} \leq 1, ~~0\leq u_{3} \leq 1, ~~0\leq u_{4} \leq 1,
$$
and the Jacobian is $\rho_{1}^2 \rho_{2} \rho_{3}$.
%This is symmetry between $\rho_2 \leftrightarrow \rho_3$, $u_1 \leftrightarrow u_2$, and $u_3 \leftrightarrow u_4$.
The change of variables
$
\rho_{1} = 1-\rho, 
 ~\rho_{2} = \rho \xi$ and
$
\rho_{3} = \rho(1-\xi)
$
gives
%\begin{equation}
$$
\int ~d\rho_{1} ~d\rho_{2} ~d\rho_{3} ~\delta(1-\sum \rho_{j})\, \rho_{1}^2 \rho_{2} \rho_{3} \cdots = \int_{0}^{1} ~d\rho \int_{0}^{1} ~d\xi ~\rho^{3} (1-\rho)^2 \xi(1-\xi)\cdots,
$$
%\end{equation}
with
$$
~~0\leq \rho \leq 1, ~~0\leq \xi \leq 1.
$$
%and Jacobian is $\rho$. The symmetry $\rho_2 \leftrightarrow \rho_3$ still remains as the symmetry $\xi \leftrightarrow (1-\xi)$.
Similar to the case of the planar diagram, $D$ and $C$ contain a common factor $\rho$. Putting ${\mathcal D} = D/\rho$ and ${\mathcal C} = C/\rho$,
delivers the final form of the integral
\begin{equation}\label{eq:integ-cross}
{I_{non-planar}} = -\int_{0}^{1} d{\rho} \int_{0}^{1}d{\xi} \int_{0}^{1} du_{1} \int_{0}^{1-u_{1}}du_{2} \int_{0}^{1} du_{3} \int_{0}^{1}du_{4}
 ~\frac{\mathcal C}{(\mathcal D-i\epsilon \mathcal C)^{3}}\,\rho(1-\rho)^2 \xi(1-\xi),
\end{equation}
where $\mathcal D$ is a quadratic in ${\bf u} =(u_1,u_2,u_3,u_4)^T,$ given by
\begin{equation}\label{eq:Dprime-cross}
\mathcal D = {\bf u}^T A {\bf u} + {\bf B}^T {\bf u} + c, 
\end{equation}
and
\begin{equation}\label{eq:Cprime-cross}
\mathcal C= \rho \xi (1-\xi) + 1 - \rho.
%= -\rho \xi^2 + \rho \xi - \rho + 1.
\end{equation}
\noindent
With the mass assignment as $m_{1}=m_{2}=m_{4}=m_{6}=m$ and $m_{3}=m_{5}=m_{7}=M (m \neq M)$ we have
%The $4\times4$ matrix $A$~ is symmetric and depends on the internal masses $m_{\ell}~~(1\le \ell\le 7)$,
%and on the kinematical variables, $s$ and $t$.
%When $m_{1}=m_{2}=m_{4}=m_{6}=m$ and $m_{3}=m_{5}=m_{7}=M, (m \neq M)$,\\
%{\footnotesize
%$$
%\begin{pmatrix}
%-m^2 \rho_{1}^2 (\rho_{2}+\rho_{3}) ~&~ (s/2- m^2)\rho_{1}^2 (\rho_{2}+\rho_{3}) ~&~ (t/2- m^2) \rho_{1}\rho_{2}\rho_{3} ~&~ (s/2+t/2- m^2) \rho_{1}\rho_{2}\rho_{3} \\
%\cdot \cdot \cdot ~&~ -m^2 \rho_{1}^2 (\rho_{2}+\rho_{3})  ~&~ (s/2+t/2- m^2) \rho_{1}\rho_{2}\rho_{3} ~&~ (t/2- m^2) \rho_{1}\rho_{2}\rho_{3} \\
%\cdot \cdot \cdot ~&~ \cdot \cdot \cdot  ~&~ - m^2 \rho_{2}^2 (\rho_{1}+\rho_{3}) ~&~ (s/2- m^2) \rho_{1}\rho_{2}\rho_{3} \\
%\cdot \cdot \cdot ~&~  \cdot \cdot \cdot ~&~ \cdot \cdot \cdot ~&~ - m^2 \rho_{3}^2 (\rho_{1}+\rho_{2}) 
%\end{pmatrix}
%$$
%}
$$A =
\begin{pmatrix}
(1-\rho)^2A_{1} ~&~ \rho\xi(1-\rho)(1-\xi)A_{2}\\
\rho\xi(1-\rho)(1-\xi)A_{2} ~&~ A_{3}\\
\end{pmatrix},
$$
\noindent
where the $2\times2$ matrices $A_{1}$, $A_{2}$ and $A_{3}$ are
$$A_{1} =
\begin{pmatrix}
-m^2~&~ s/2-m^2\\
s/2-m^2~&~ -m^2\\
\end{pmatrix},
\qquad
A_{2} =
\begin{pmatrix}
t/2-m^2~&~ s/2+t/2-m^2\\
s/2+t/2-m^2~&~t/2-m^2 \\
\end{pmatrix},
$$

$$A_{3} =
\begin{pmatrix}
-m^2\rho\xi^2(1-\rho\xi)~&~ (-s/2+m^2)\rho(1-\rho)\xi(1-\xi)\\
(-s/2+m^2)\rho(1-\rho)\xi(1-\xi)~&~ -m^2\rho(1-\xi)^2(1-\rho(1-\xi))\\
\end{pmatrix}.
$$
\noindent
The vector $\bf B$ is given by
$${\bf{B}} =
\begin{pmatrix}
-t \rho\xi(1-\rho)(1-\xi) + M^2 (1-\rho){\mathcal C}\\
-t \rho\xi(1-\rho)(1-\xi) + M^2 (1-\rho){\mathcal C}\\
-t \rho\xi(1-\rho)(1-\xi) + M^2 \rho\xi{\mathcal C}\\
-t \rho\xi(1-\rho)(1-\xi) + M^2 \rho(1-\xi){\mathcal C}
\end{pmatrix},
$$
\noindent
and the scalar $c$ is 
\begin{equation}
c = t\rho(1-\rho)\xi(1-\xi) -M^2 {\mathcal C}.\nonumber
\end{equation}

Note the similarity between the Eqs. (\ref{eq:integ-ladder}) and (\ref{eq:integ-cross}) of the planar and the non-planar integral, respectively, obtained via the transformations in Sections~\ref{subsec:tlbl} and ~\ref{subsec:tlbc}. 
The transformed integrand functions (of both 6-dimensional integrals) involve a function $\mathcal{D}$ in the denominator, which is a quadratic in the variables $u_1$, $u_2$, $u_3$ and $u_4$. 
This form of the integrand will further lend itself to the reduction method of Section~\ref{subsec:reduction} (which will be used for a comparison of the numerical results). 
As an aside, the form of the 4-dimensional integral in $u_1$, $u_2$, $u_3$, $u_4$ also resembles that of the one-loop pentagon integral, {\it e.g.}, in \cite{jocs11}. 
%-----------------------------------------------------------------------

%%%%%%%%% Section 3  %%%%%%%%%%
\section{Numerical techniques}\label{sec:tech}
We introduce the {\it Direct Computation Method} (DCM), based on a combination of numerical integration, and extrapolation on a sequence of integrals.
DCM comprises the following three steps:
\begin{enumerate}
\item
Let $\epsilon$ in Eq.\,\eqref{eq:scalar-integral} be a finite value determined by a (scaled) geometric sequence 
\begin{equation}
\epsilon = \epsilon_l = \epsilon_0/(A_c)^l,l=0,1,\cdots, 
\label{eps-sequence}
\end{equation}
for a constant $\epsilon_0$ and base $0 < 1/A_c < 1.$ 
\item
Evaluate the multi-dimensional integral $I$ of Eq.\,\eqref{eq:scalar-integral} numerically. In view of the finite $\epsilon_l$ we obtain a finite value for the integral corresponding to each $l$. Thus a sequence of $I(\epsilon_l)$, $l=0,1,2,\cdots$ is generated.
\item
Extrapolate the sequence $I(\epsilon_l)$ to the limit as $\epsilon_l \rightarrow 0$ with the purpose of calculating 
$I$ as $\lim_{\epsilon\rightarrow 0} I(\epsilon).$
\end{enumerate}
If $D$ does not vanish within the integration region, we can ignore $\epsilon$ and no extrapolation is needed as $I = I(\epsilon)\mid_{\epsilon=0}$. 

For multi-dimensional integration we make use of the {\tt DQAGE} routine in the {\tt QUADPACK}~\cite{quadpack} package.
{\tt DQAGE} uses a variant of Gaussian quadrature, where the sampling points are given by a Gauss-Kronrod rule pair 
in each subinterval. The Gauss rule with $\nu$ points has polynomial degree of accuracy $d_\nu = 2\nu-1;$ i.e., it is exact for 
polynomials of degree $d = 0,\cdots,d_\nu$ and not for all polynomials of degree $d_\nu+1.$ The corresponding Kronrod rule
re-uses the abscissas of the Gauss rule and adds $\nu+1$ points interlacing with those of the Gauss rule. The Kronrod rule with $2\nu+1$ points has polynomial degree $3\nu+1$ if this number is odd (for $\nu$ even), and otherwise $3\nu+2$
(for $\nu$ odd).

On input for {\tt DQAGE}, the user selects one of six Gauss-Kronrod pairs, with 15, 21, 31, 41, 51 or 61 points, via the input parameter 
${\tt key} = 1,2,3,4,5, \mbox{ or } 6,$ respectively. 
The rule pair produces the Kronrod rule value as the integral approximation, together with an estimate of the absolute 
error on each subinterval (which is based on the difference between the Gauss and the Kronrod result on the subinterval).
This allows the selection of that subinterval with the largest estimated error, as the next interval to be subdivided 
in successive steps of the adaptive partitioning strategy of {\tt DQAGE}. 
The user imposes a bound on the number of subdivisions via the input parameter ${\tt limit}.$
As a result of the adaptive partitioning, the algorithm subdivides intensively around singularities, so that
hot spots emerge where singularities or other irregular integrand behavior occur within the integration interval.
For multi-dimensional integration we apply {\tt DQAGE} in a repeated (iterated) quadrature for successive coordinate directions~\cite{iterate}.

In DCM, the accuracy of the result depends on that of the calculated sequence of integrals $I(\epsilon_l), l=0,1,\cdots.$ Since the integration error
affects the accuracy of the extrapolation, we want to compute the integrals $I(\epsilon_l)$ to at least an order of magnitude more accuracy than that expected for the final result. On the other hand, the CPU time is directly related to the accuracy 
requirement.
Thus, adequate values need to be specified for the input parameters ${\tt key}$ and ${\tt limit}$ of the {\tt DQAGE} 
routine in each dimension, to control the overall work and the termination of the algorithm.
For the computation of two-loop box integrals, we find that ${\tt key}=1\hbox{ or }2$ and ${\tt limit}=10 \sim 30$ are suitable values.

We use Wynn's $\varepsilon$-algorithm\cite{shanks,wynn} for the extrapolation, which
works efficiently under fairly general conditions, even for very slowly convergent sequences or series.
The $\varepsilon$ algorithm is applied to the sequence $I(\epsilon_l), l=0,1,\cdots$ obtained by multi-dimensional 
integration. We define the table elements $a(l,k)$ of the extrapolation table with initial values
\begin{equation}
a(l,-1)=0, \qquad a(l,0)=I(\epsilon_l), \qquad l=0,1,\cdots.
\end{equation}
The element $a(l,k+1)$ is obtained from $a(l,k)$ and $a(l,k-1)$ by the following recurrence relation:
\begin{equation}
a(l,k+1)=a(l+1,k-1)+\frac{1}{a(l+1,k)-a(l,k)}, \qquad l=0,1,\cdots.
\end{equation}
Whilst the $a(l,k)$'s with odd $k$ are meant to store temporary numbers, the $a(l,k)$'s with even $k$ give extrapolated estimates.

We use the $\varepsilon$-algorithm code from the {\tt QUADPACK}~\cite{quadpack} package.
With each new $I(\epsilon_l),$ a new lower diagonal can be added to the extrapolation table.
At each iteration only the last two lower diagonals need to be stored for this computation.
Along with each new table element $a(m,n)$ where $n$ is even, an error estimate is calculated based on differences
with its neighboring elements.
In a converging table, the even-numbered columns as well as the diagonals converge to the limit 
$\lim_{\epsilon\rightarrow 0} I(\epsilon)=I$ (barring roundoff). 
Among the even-column indexed table elements along newly computed lower diagonal, 
the $\varepsilon$-algorithm code selects the "best" $a(m,n)$ (with the least error estimate).
The CPU time for the extrapolation is negligible compared to that of the integration.

We further have to use some heuristics for the computation of the extrapolated sequence.
The acceleration constant $A_c$ in~\eqref{eps-sequence} can usually be set to 2. 
In cases where the integration is very difficult for decreasing $\epsilon_l,$ a smaller value of $A_c$ is used, {\it e.g.}, 
$A_c=1.3$ or $1.2,$ yielding a sequence of $\epsilon_l, l = 0, 1,\cdots$ which decreases more slowly.
To determine the initial value of the geometric sequence, we assign $\epsilon_0$ depending on the squared mass appearing in function $D$. 
%($M^2$) which enters via the scalar $c$ in the $\mathcal{D}$ function~\eqref{eq:Dprime-ladder} and~\eqref{eq:Dprime-cross}. of the loop integrand denominator. 
We parametrize $\epsilon_0$ in the form $\epsilon_0=A_c^\gamma,$ 
where the parameter $\gamma$ can be adjusted.
The choice of these parameters influence the accuracy of the result. 

For the two-loop box integral computations reported in the next section, we found $A_c=1.2$ and $\gamma$ around 40 
to be adequate values. 
The accuracy achieved is restricted by the actual CPU time needed. If the computation time is excessive, 
we have to accept less accurate results. This happens, for example, when $s$ is much greater than $10m^2$. 
Thus the accuracy is different from point to point in the plots shown below. 
All the computations are done in double precision arithmetic.
%-----------------------------------------------------------------------

%%%%%%%%% Section 4  %%%%%%%%%%
\section{Numerical results}\label{sec:results}
According to the prescription of DCM in the previous section, we evaluate both $I_{planar}$ and $I_{non-planar}$ given by Eqs. (\,\ref{eq:integ-ladder}) and~\eqref{eq:integ-cross}, respectively. In both cases the kinematical variable $s$ is varied but $t$ is fixed at $t=-10000{\rm GeV}^2$
 throughout the computations. We introduce the dimensionless variable
\begin{equation}
f_s=\frac{s}{m^2}.
\label{fs}
\end{equation} 
For the mass parameters we set $m=50 {~\rm GeV}$ and $M=90 {~\rm GeV}$.

\subsection{Planar diagram}\label{subsec:ladder-result}
\begin{figure}
\centering
\includegraphics[width=0.7\linewidth]{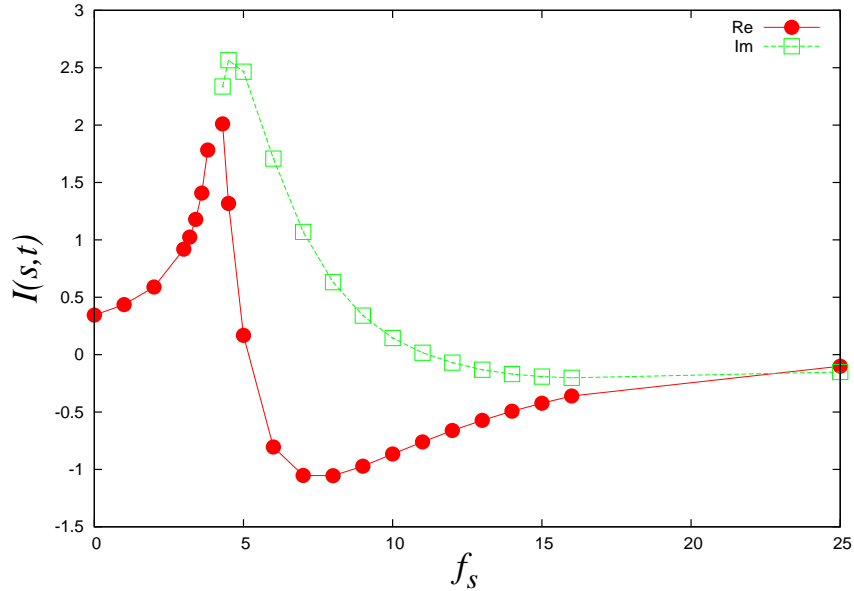}
\caption{Numerical results of $\Re(I_{planar})$  and $\Im(I_{planar})$ in units of $10^{-12}$ ${\rm GeV}^{-6}$ for $0.0 \le f_{s} \le 25.0$ and $t=-10000.0 {\rm GeV}^2$. Plotted points are the real part (bullets) and the imaginary part (squares).}
\label{fig:tbl-fsp}
\end{figure}

In previous work~\cite{jocs11} we presented results of the real part integral in the physical region, $4.5 \leq f_s \leq 25.0$.
Here we evaluate the integral in the region $0.0 \leq f_s \leq 25.0$ for the real part and the imaginary part. 
The results are depicted in Fig.~\ref{fig:tbl-fsp} where the data points represent the integral values
and the lines merely connect the points as a guide for the eyes. The $s$-channel threshold starts at $f_s=4.0$ corresponding 
to $s=4m^2$. 
For example, we set $\epsilon_0 = 1.2^{45}$,  ${\tt key}=1$ and ${\tt limit}=10$ in all dimensions for the real part at $f_s=10.0$ 
and it took 8.5 days to obtain the result with enough accuracy as 0.01\% using a system with Intel Xeon CPU E5430 @ 2.66GHz.
%1.2^{45}=3657.2

\subsection{Non-planar diagram}\label{subsec:cross-result}
\begin{figure}
\centering
\includegraphics[width=1.25\linewidth,angle=90]{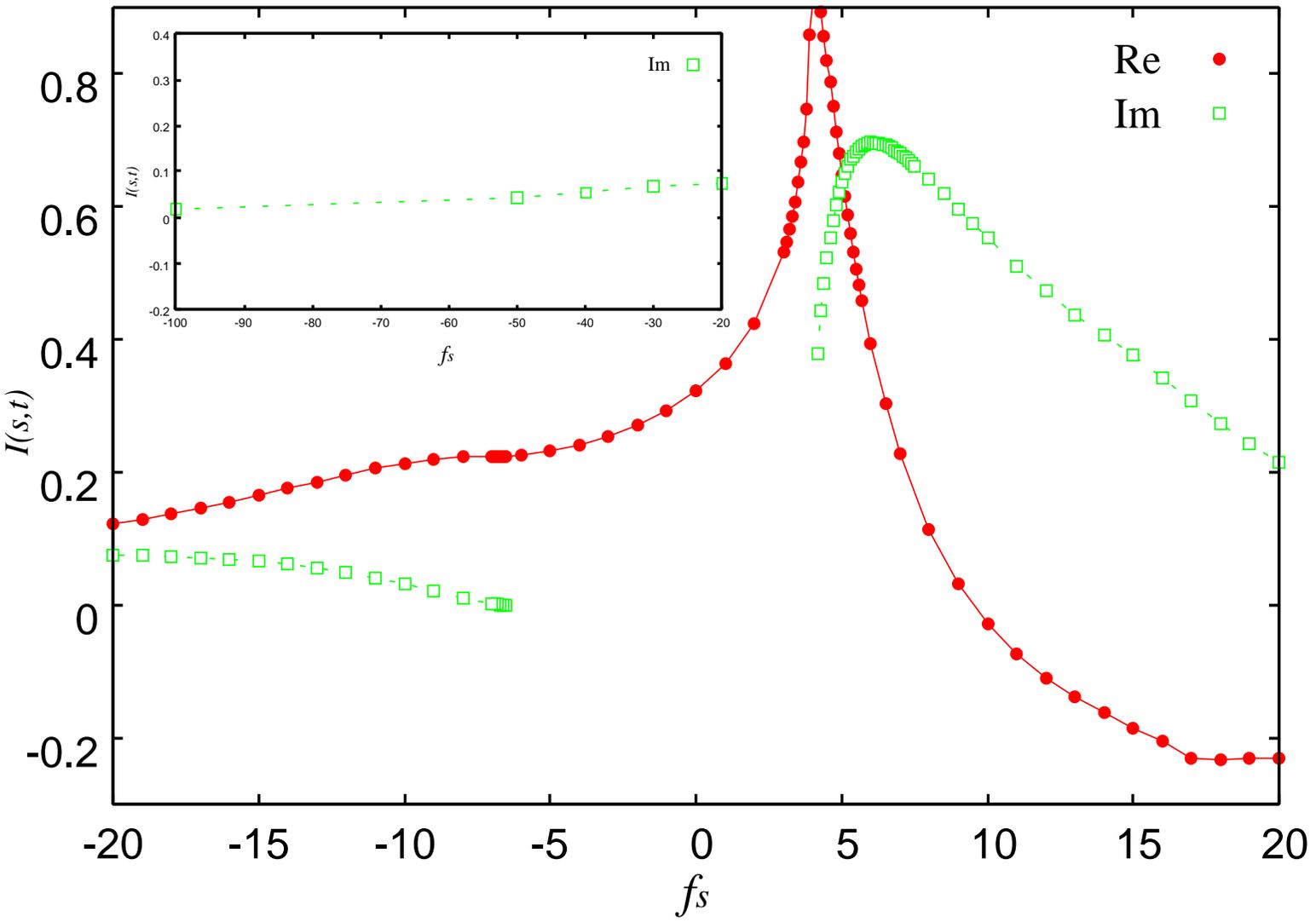}
\caption{Numerical results of $\Re(I_{non-planar})$ and $\Im(I_{non-planar})$ for $-20.0 \le f_{s}\le 20.0$ and of $\Im(I_{non-planar})$ for $-100.0 \le f_{s}\le -20.0$ with $t=-10000.0 {\rm GeV}^2$ in units of $10^{-12}$ ${\rm GeV}^{-6}$. Plotted points are the real part (bullets) and the imaginary part (squares).}
\label{fig:tbc}
\end{figure}

Fig.~\ref{fig:tbc} shows the results for $-20.0 \le f_s \le 20.0$.
Different from the planar case, it is known that $I_{non-planar}$ has two cuts; one starts from the normal $s$-channel threshold, $s=4m^2$, 
and the other from $s=-t-M^2-4mM$ to $s=-\infty$. The latter corresponds to the $u$-channel threshold at $u=(M+2m)^2$. 
These correspond to $f_{s}=4.0$ and $f_{s}=-6.44$, respectively. 
In Fig.~\ref{fig:tbc} we also show some results of the imaginary part in the range $-100.0 \le f_s \le -20.0$. In this region the imaginary part is small but its contribution is not 
negligible when it is put in the dispersion integral~\eqref{eq:disp-all} of Section~\ref{subsection:consistency}.

%The imaginary part becomes smaller when $f_s$ takes negative and smaller values. 
%The imaginary part in Fig.\ref{fig:tbc-large-fsm} in the $f_s$ region of Fig.\ref{fig:tbc-large-fsm} is small but still necessary to compare the results with the dispersion relation in 

%\begin{figure}
%\centering
%\includegraphics[width=0.7\linewidth]{6d.cross.DQ.fsm.20110828.eps}
%\caption{Numerical results of $\Re(I_{non-planar})$ and $\Im(I_{non-planar})$ in units of $10^{-12}$ ${\rm GeV}^{-6}$ for $-20.0 \le f_{s}\le 0.0$. Plotted points are the real part (bullets) and the imaginary part (squares).}
%\label{fig:tbc-fsm}
%\end{figure}

%\begin{figure}
%\centering
%\includegraphics[width=0.7\linewidth]{6d.cross.DQ.large-fsm.20110828.eps}
%\caption{Numerical results of $\Im(I_{non-planar})$ in units of $10^{-12}$ ${\rm GeV}^{-6}$ for $-100.0 \le f_{s}\le -20.0$. 
%Calculated points are depicted by squares.}
%\label{fig:tbc-large-fsm}
%\end{figure}

In the physical region the computation time tends to be longer for larger $f_s.$ This applies to both the planar and the
non-planar diagram. For example the time required to obtain the real part of the non-planar box integral with enough accuracy as 0.003\% at  $f_s=10.0$ (with $\epsilon_0 = 1.2^{40}$, ${\tt key}=2$ 
in all dimensions, and ${\tt limit}=10,20,20,10,10,10$ in consecutive dimensions) is about a week using a system with Intel Xeon CPU X5365 @ 3.16GHz. For much greater $f_s$ it may become more difficult to get an answer in a practical time. However, this computation 
time is measured using a single CPU. 
It can potentially be shortened by applying parallel computing techniques
on (possibly distributed) multi-core processors~\cite{acat08,iciam11,ccp11}.
%-----------------------------------------------------------------------

%%%%%%%%% Section 5  %%%%%%%%%%
\section{Validation of the results}\label{sec:quality}
After obtaining answers by the numerical computation, the most important issue is how to confirm 
that the results are correct and reliable. It would be most desirable to have answers available by independent methods.
For example, we computed the result $0.10364072096 \pm (0.315 \times 10^{-7})$ for the planar integral, with
$s=t=1$ $\rm{GeV}^2$ and $m_i^2=1$ ${\rm GeV}^2 ~(1 \le i \le 7),$ $p^2_i = 1$ ${\rm GeV}^2 ~(1 \le i \le 4).$ We were able to compare this to the value $0.1036407209893$ evaluated by the program {\tt SYS}~\cite{laporta} and found good agreement. 
With the same values of the kinematical variables we obtained the non-planar integral as 
$0.08535139 \pm (0.105 \times 10^{-7}),$ but no result was available by {\tt SYS}. 
This comparison demonstrates that the expressions of the functions $C$ and $D$ for the planar diagram are correct and 
that {\tt DQAGE} works as expected.
In these examples, $D$ does not vanish within the integration region; thus we do not need extrapolation.
The CPU time required for both the planar and the non-planar diagram is less than 2 {\it min.} using a Xeon CPU X5365 @ 3.16GHz.

Below we outline an integration method based on reduction formulas, and explain how to use it for a consistency check.

%-----------------------------------------------------------------------
\subsection{Comparison with the reduction method}\label{subsec:reduction}
Consider a quadratic form ${\mathcal D}$ in $N$ variables,
${\bf u}=(u_1,\cdots,u_N),$
\begin{equation}
{\mathcal D}={\bf u}^T A {\bf u} + {\bf B}^T{\bf u} + c,
\end{equation}
where $A$ is an $N$-dimensional symmetric matrix $A=(A_{ij})$, ${\bf B}$ is an $N$-dimensional vector ${\bf B}^T=(B_1,\cdots,B_N)$,
with constant coefficients. Here $c$ contains not only a real part but also $-i\epsilon{\mathcal C}$ 
with an infinitesimal $\epsilon$. Assuming $A$ is invertible, we define the vector 
\begin{equation}
{\bf X}^T =2{\bf u}^T+ {\bf B}^T A^{-1}.
\end{equation}
Then we have
\begin{equation}\label{eq:eqtwofive}
{\bf X}^T (\nabla {\mathcal D})=4({\mathcal D}-c)+{\bf B}^TA^{-1}{\bf B}=4{\mathcal D}+\Delta_N,
\end{equation}
with $\nabla^T=(\partial/\partial u_1,\cdots,\partial/\partial u_N)$. Here  $\Delta_N$ is defined as
\begin{equation}
\Delta_N={\bf B}^TA^{-1}{\bf B}-4c.
\end{equation}
%It is assumed that $\Delta_N$ does not vanish. 
We divide Eq.\,\eqref{eq:eqtwofive} by ${\mathcal D}^{\alpha+1}$ where
$\alpha \ge 0$ is an arbitrary number. Using the relation
\begin{equation}
\nabla^T\left(\frac{{\bf X}}{{\mathcal D}^{\alpha}}\right)
=  \frac{2N}{{\mathcal D}^{\alpha}}-\alpha  \frac{{\bf X}^T \nabla {\mathcal D}}{{\mathcal D}^{\alpha+1}}
\end{equation}
we obtain the following reduction formula
\begin{equation}
\frac{\Delta_N}{{\mathcal D}^{\alpha+1}}=\frac{-4+2N/\alpha}{{\mathcal D}^{\alpha}}
-\frac{1}{\alpha}\nabla^T\left(\frac{{\bf X}}{{\mathcal D}^{\alpha}}\right), ~~\mbox{ for } \alpha > 0.
\end{equation}
It should be noted  that the power of the denominator in the right-hand side is decreased by one, compared to the left-hand side,
that is, the singular behavior is softened. When $\alpha=0$ we find 
\begin{equation}
\frac{\Delta_N}{{\mathcal D}}=-4-2N\log {\mathcal D}
+\nabla^T\left({\bf X} \log {\mathcal D} \right).
\end{equation}
When a polynomial in ${\bf u}$, $f({\bf u})\neq 1$, occurs in the numerator of the left-hand side, the formulas are 
generalized to 
\begin{equation}
\frac{f \Delta_N}{{\mathcal D}^{\alpha+1}}
=\frac{-4f+\nabla^T(f{\bf X})/\alpha}{{\mathcal D}^{\alpha}}
-\frac{1}{\alpha}\nabla^T\left(\frac{f {\bf X}}{{\mathcal D}^{\alpha}}\right),~~\mbox{ for } \alpha > 0
\end{equation}
and
\begin{equation}
\frac{f \Delta_N}{{\mathcal D}}=-4f-\nabla^T(f{\bf X})\log {\mathcal D}
+\nabla^T\left(f{\bf X} \log {\mathcal D} \right).
\end{equation}
We apply the formula to the functions ${\mathcal D}$ given in Eqs.~\eqref{eq:Dprime-ladder} and~\eqref{eq:Dprime-cross},
which are quadratics in $u_1,\cdots, u_4$. Since we have $N=2\alpha ~(N=4,\alpha=2)$ in both cases, we find a simpler formula
\begin{equation}\label{eq:D4}
\frac{\Delta_4}{{\mathcal D}^3} = -\frac{1}{2}\nabla^T\left(\frac{\bf X}{{\mathcal D}^2}\right),
\end{equation}
with $\Delta_4=\Delta_4(\rho, \xi)$.
By integrating we have
\begin{equation}\label{eq:D4integral}
\int d({\bf u}) \frac{1}{{\mathcal D}^{3}}= -\frac{1}{2\Delta_4}\int d({\bf u})\nabla^T\left(\frac{{\bf X}}{{\mathcal D}^2}\right), \qquad d({\bf u})=\prod_{j=1}^4du_j.
\end{equation}
\noindent
In Eqs.(\ref{eq:integ-ladder}) and (\ref{eq:integ-cross}), the above expression is integrated over $\rho$ and $\xi$.  
%Here $\Delta_4=0$ can occur in $\rho\hbox{-}\xi$ space.
Here $\Delta_4=0$ can occur in $\rho$-$\xi$ space, and is regularized numerically by setting the integrand to zero in the vicinity of this anomaly.
%This is not the real singularity and can be regularized numerically, for example, avoiding the region where $|\Delta_4|$ is very small.

The right-hand side is immediately integrated once. Applying the reduction repeatedly to the form in the right-hand 
side, we see that the original integral is finally replaced by a sum of integrals of functions involving logarithms. Thus the severity of the 
integrand singularity is reduced, which allows performing the integration even with Monte Carlo routines. Note that this procedure generally creates lengthy 
expressions. The imaginary part results from the logarithms; let $R$ be a positive number and let $z=-R\pm i\epsilon$,
then $\log z=\log(-R\pm i\epsilon)=\log R\pm i\pi$. We refer to this integration method as the {\it Reduction Method} (RM). 
We computed the two-loop box integrals by using {\tt BASES}~\cite{bases}. The real part of the planar diagram integral in the physical 
region, shown in~\cite{jocs11}, is in good agreement with the results by DCM. On the other hand, for the imaginary 
part, the Monte Carlo integration failed to convergence satisfactorily. 

\begin{figure}
\centering
\includegraphics[width=0.7\linewidth]{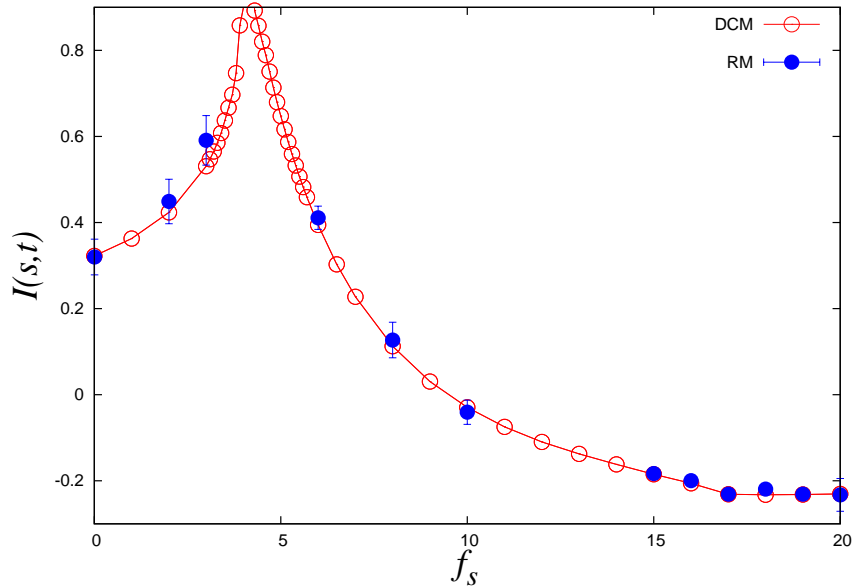}
\caption{Numerical results of $\Re(I_{non-planar})$ in units of $10^{-12}$ ${\rm GeV}^{-6}$ for $0.0 \le f_{s} \le 20.0.$ Values calculated by DCM and those by RM are shown by circles and bullets with error bars, respectively.}
\label{fig:tbc-reduction}
\end{figure}

In Fig.~\ref{fig:tbc-reduction} we show the real part of the non-planar case obtained by the reduction formulas. Agreement with 
the results by DCM is poor around the threshold $f_s=4$ in view of poor convergence of the integration by RM.
%This might come from the numerical regularization of $\Delta_4=0$ referred below Eq.(\ref{eq:D4integral}).
This may be caused by the numerical regularization in the vicinity of $\Delta_4 = 0$ as mentioned above.
%When $\Delta_4$ becomes very small, it causes instabilities and we take a trick to aboid it numerically.
%However, the convergence improves significantly as $f_s$ increases.

%-----------------------------------------------------------------------
\subsection{Consistency check using dispersion relation}\label{subsection:consistency}
The dispersion relation provides a good tool for a consistency check. 
Based on the observation that $I(s)$ can be regarded as an analytic function in the complex $s$-plane, 
the real part and the imaginary part, $\Re(I(s))$ and $\Im(I(s)),$ of $I(s)$ satisfy the {\it dispersion relation},
\begin{equation}\label{eq:disp-all}
\Re(I(s))=\frac{1}{\pi}\,{\mathtt P}\hspace*{-1mm}\int_{-\infty}^{{+\infty}} \frac{\Im(I(s^\prime))}{s-s^\prime}ds^\prime,
%\Re(I(s))=\frac{1}{\pi}\left({\mathtt P}\int_{-\infty}^{s_{0}^\prime} \frac{\Im(I(s^\prime))}{s-s^\prime}ds^\prime + {\mathtt P}\int_{s_{0}}^{\infty} \frac{\Im(I(s^\prime))}{s-s^\prime}ds^\prime \right),
\end{equation}
\noindent
where ${\tt P}$ denotes principal value integral. Recall that $I(s)$ is real in some region of $s$ and accordingly 
$\Im(I(s))$ vanishes there. This integral relation, which is the consequence of the analyticity of $I(s)$, implies the real
part can be estimated from the imaginary part. DCM computes the real and the imaginary part independently, as 
they are given by separate integrals. However, the dispersion relation indicates that both parts are not independent. 
They should be consistent with the relation of Eq.\,\eqref{eq:disp-all}.

In order to show how this relation works we consider the two-loop non-planar box as an example. 
In this case the integral can be written as
\begin{equation}\label{eq:disp}
\Re(I(s))=\frac{1}{\pi}\,\left({\mathtt P}\hspace*{-1mm}\int_{-\infty}^{s_{0}^\prime} \frac{\Im(I(s^\prime))}{s-s^\prime}ds^\prime + {\mathtt P}\hspace*{-1mm}\int_{s_{0}}^{\infty} \frac{\Im(I(s^\prime))}{s-s^\prime}ds^\prime \right),
\end{equation}
\noindent
where $s_{0}=4m^2$ and $s_{0}^\prime=-t-M^2-4mM$ are the threshold in the $s$-channel and in the $u$-channel, respectively,
as described in Section~\ref{subsec:cross-result}. 

\begin{figure}
\centering
\includegraphics[width=0.7\linewidth]{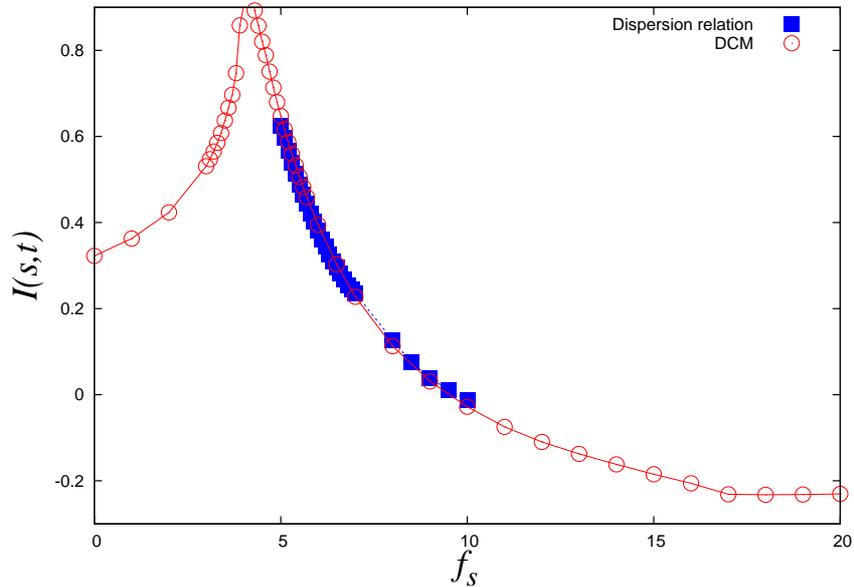}
\caption{Numerical results of $\Re(I_{non-planar})$ in units of $10^{-12}$ ${\rm GeV}^{-6}$ for $0.0\le f_s\le20.0.$ 
Values calculated by DCM and those by the {\it dispersion relation} are shown by circles and squares, respectively.}
\label{fig:tbc-disp}
\end{figure}

For the principal value integral computation we used the trapezoidal rule, assuming that $\Im(I(s)) = 0$ far away from the origin,
for $f_s\le -100.0$ and $f_s\ge50.0.$ Values of $\Re(I(s))$ resulting from this computation are plotted in Fig.~\ref{fig:tbc-disp} for $0.0 \leq f_s \leq 10.0.$ 
The results show good agreement with those by DCM. Thus the relation of Eq.\,\eqref{eq:disp-all} enables a consistency check 
for the answers produced by DCM.

%-----------------------------------------------------------------------

%%%%%%%%% Section 6  %%%%%%%%%%
\section{Conclusions}\label{sec:summary}

In this paper we calculated the scalar integrals of two-loop planar and non-planar box diagrams involving massive particles.
We introduced the {\it Direct Computation Method} (DCM) for the evaluation. The novel idea in DCM is that the $\epsilon$ value in the propagators 
is treated numerically as a finite number, not as an infinitesimal value. In view of the finite $\epsilon,$ the integrand of the loop integral is 
no longer singular. The integration can be carried out numerically for both the real part and the imaginary part. Consecutive integrations,
for each $\epsilon_l,$ produce a sequence of integrals $I(\epsilon_l), l = 0,1,\cdots,$ which are supplied to the extrapolation procedure. 
A numerical answer for the loop integral results in the limit as $\epsilon_l$ tends to 0.

Since DCM does not impose restrictions on the values of mass parameters, the method is valid when masses are complex~\cite{acat08}.
For this case we can put $\epsilon = 0$ where no extrapolation is needed in the same manner as in the non-physical region.
This flexibility is remarkable and useful for the calculation of cross-sections where decaying particles are involved. 

In order to check our evaluations we compared the results with those obtained by other methods~\cite{laporta}, including {\it Reduction Method} (RM). 
Comparisons of the results have shown satisfactory agreement. The examination of the dispersion relation
has lead to a consistency check between the real and the imaginary part of the integral. Thus we have established various
ways to confirm the results.

Some issues linked with the implementation of parameters, {\it e.g.}, the choice of $\epsilon_l$ values, need to be solved heuristically. Furthermore, 
in some regions of the kinematical variables, DCM requires very long 
CPU times to obtain reasonable accuracy. 
It may be possible to tackle the CPU time problem by utilizing recent developments in computer resources and parallel computing technologies~\cite{acat08,iciam11,ccp11}. 
Throughout this paper we use double precision arithmetic, but quadruple or extended precision may be needed for some mass configurations, including a small fictitious mass $\lambda$ to regularize infrared divergent integrals~\cite{acat07}. 
This can be incorporated in dedicated program packages~\cite{acat05,fujiwara}.

For a specified high-energy reaction, all the necessary two-loop diagrams can be generated automatically using
the GRACE system~\cite{grace}. The next stage, which involves the automatic generation of amplitudes ({\it i.e.}, the integrands of loop-integrals), 
would be manageable in view of the experience we gained in handling tree and one-loop processes~\cite{grace}. 
Thus the only component which needs further development for the construction of an automatic computation system for two-loop reactions is a robust loop integral evaluation system.

Concerning the further development of DCM we need to test integrals for various mass configurations different from 
those in this paper, particularly, infrared divergent integrals by the prescription using a fictitious mass. We also need to examine loop integrals with a
non-trivial numerator, and explore a systematic treatment of ultra-violet divergence. 
From a technical point of view, reducing CPU time and automatic tuning of the integration parameters should be included.
After completion of these studies, 
we expect that DCM will play an important  role in constructing automatic computation systems for  higher-order corrections. 

~\par
{\large{\bf{Acknowledgements}}}\\

We wish to thank Prof. T.Kaneko for valuable discussions and comments. 
This work was supported in part by the Grant-in-Aid (No.20340063 and No.23540328) of JSPS and by the CPIS program of Sokendai.

%%%%%%%%% Appendix A  %%%%%%%%%%
%\\
%\par
%{\large{\bf{Appendix}}}
\appendix
\section{Construction of the functions $D$ and $C$}\label{app:DandC}
For a given diagram, the explicit form of the functions $D$ and $C$ is determined by the following steps~\cite{tiktopoulos,eden,nakanishi}.
\begin{itemize}
\item{Step 1.
\begin{enumerate}
\item{Assign the parameter ${x_i}$ to the $i$-th internal line. 
The parameters $\{x_i\}$ satisfy $\displaystyle{\sum_{i=1}^N x_i = 1}$.
}
\item{Define $L$ topologically independent loops 
and label them as $a=1, \cdots ,L$.
The loop momentum $l_a$
flows through the $a$-th loop in its own direction.
}

\item{External momenta $p_j, j=1, \cdots ,K$ are presumed to 
enter the diagram inward. Here $K$ is the number of external lines.
We let $p_j$ flow through the diagram while
respecting the momentum conservation at each vertex.
A simple example is where each $p_j, j=1, \cdots ,K-1$
flows through the diagram along a continuous path, to reach the
vertex where $p_K$ enters. In this case the momentum conservation
is trivial as $\sum_{j=1}^K p_j =0$.
}

\item{
Each internal line has its direction and the momentum $k_i$ for the $i$-th internal line is defined along this direction.
It can be expressed by a linear combination of the $l_a$ and $p_j$ as
$$ k_i = \sum_{a=1}^L \sigma_a^i l_a + \sum_{j=1}^K \tilde{\sigma}_j^i p_j $$
where 
$$
\sigma_a^i= \left\{\begin{array}{cl}
1 &  l_a\ \mathrm{flows}\ \mathrm{along}\ \mathrm{the}\ 
i\mathrm{\hbox{-}th\ internal\ line\ parallel\ to\ its\ direction} \\
-1 &  l_a\ \mathrm{flows}\ \mathrm{along}\ \mathrm{the}\  
i\mathrm{\hbox{-}th\ internal\ line\ anti\hbox{-}parallel\ to\ its\ direction} \\
0 & \mathrm{(otherwise)}
\end{array}\right. $$
and $ \tilde{\sigma}_j^i $ can be defined in a similar manner
for $p_j$.
We define $p_{ext,i}=\sum_{j=1}^K \tilde{\sigma}_j^i p_j $ 
for the $i$-th internal line.
}
\item{It should be noted that, even though the choice of
the $L$-loops, the selection of loop-momenta $l_a$, 
and the flow of external momenta are not unique,
the final result is the same for any choice.}
\end{enumerate}
}
\item{Step 2.\\
We construct an $L \times L$ symmetric matrix $\mathsf A$, an $L$-vector ${\mathsf B}$ and a scalar $\mathsf c$.
$$
\mathsf A_{ab}=\sum_{i=1}^N \sigma_a^i\sigma_b^i x_i,\qquad
\mathsf B_{a}=\sum_{i=1}^N \sigma_a^i x_i p_{ext,i},\qquad
\mathsf c=\sum_{i=1}^N x_i(p_{ext,i}^2-m_i^2).
$$
}
\item{Step 3. \\
The functions $C$ and $D$ are obtained by
$$
C={\rm det}({\mathsf A}), \qquad {\rm and} \qquad
D = -{\rm det}  
\left(
\begin{array}{cc}
{{\mathsf A}} & {{\mathsf B}} \\
{{\mathsf B}}^T & {{\mathsf c}}
\end{array}
\right).
%\begin{vmatrix}
%{A} & {B} \\ 
%{B}^T & {c} 
%\end{vmatrix}.
$$
$D$ is a homogeneous polynomial of degree $L+1,$ and
$C$ is a homogeneous polynomial of degree $L$ in $x_i$.
}
\end{itemize}

\section{Two-loop box diagrams}\label{app:tlb}
\begin{center}
\begin{figure}[htb]
\subfigure[] {
\includegraphics[width=0.45\linewidth]{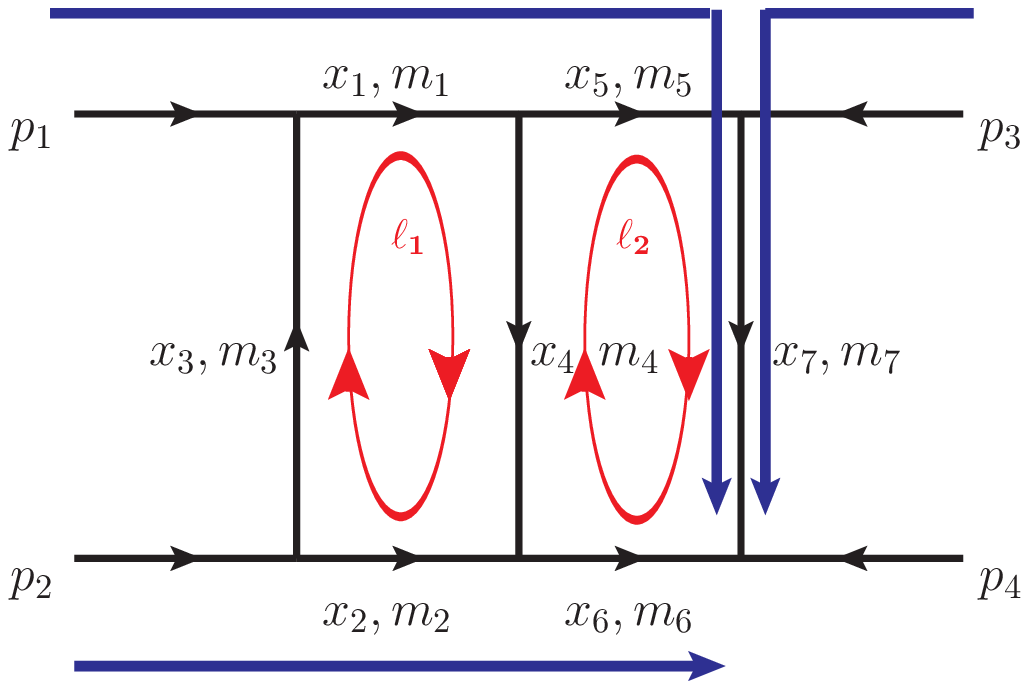}\label{fig:tlb-a-appen}}
\subfigure[] {
\includegraphics[width=0.45\linewidth]{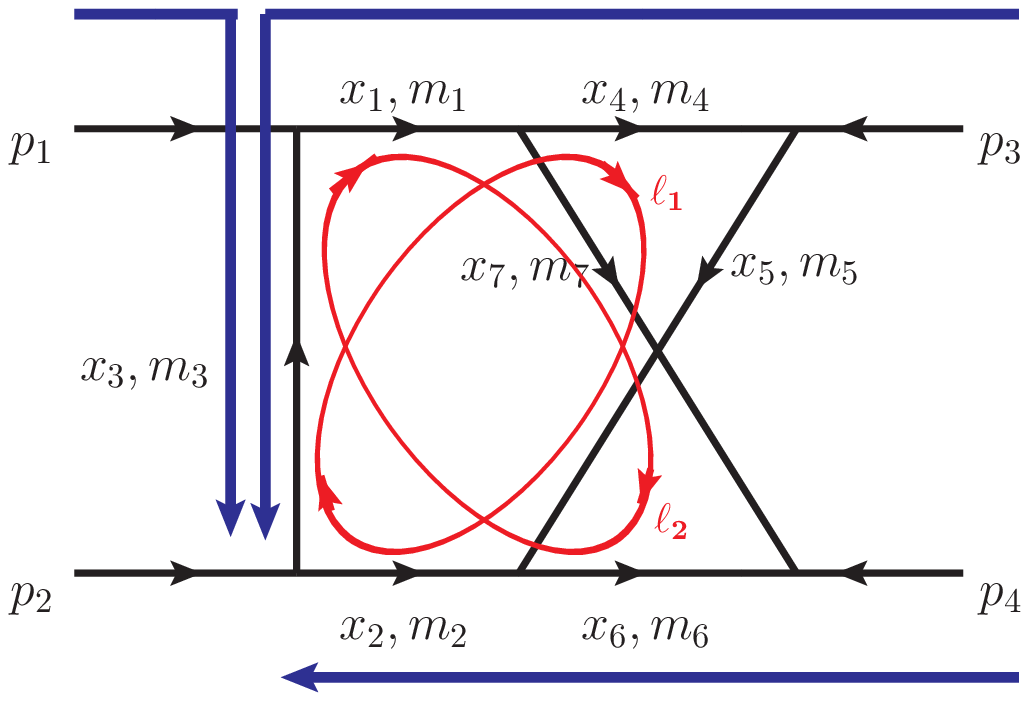}\label{fig:tlb-b-appen}}
\caption{The quantities in \ref{app:tlb} are obtained from the configuration shown in the figure for (a) two-loop planar box and (b) two-loop non-planar box. The arrow on each internal line defines its direction(Step 1.4). Red lines and blue lines show the flow of loop momenta($l_a$, Step 1.2) and that of external momenta($p_j$, Step 1.3), respectively.}
\end{figure}
\end{center}

Following these prescriptions, $D$ and $C$ (Eqs. (\ref{eq:D-ladder}), (\ref{eq:C-ladder})) for the two-loop planar diagram (Fig.~\ref{fig:tlb-a-appen}) are obtained from
\begin{eqnarray*}
&&{{\mathsf A}}_{11}=x_1+x_2+x_3+x_4, \qquad{{\mathsf A}}_{12}={{\mathsf A}}_{21}=-x_4, \qquad{{\mathsf A}}_{22}=x_4+x_5+x_6+x_7,\\
&&{{\mathsf B}}_1=x_1p_1 - x_2p_2, \qquad{{\mathsf B}}_2=x_5p_1 - x_6p_2 +x_7(p_1+p_3),\\
&&{\mathsf c}=x_1(p_1^2-m_1^2) + x_5(p_1^2-m_5^2) + x_2(p_2^2-m_2^2) + x_6(p_2^2-m_6^2) \\
&& \qquad\qquad+x_7((p_1+p_3)^2-m_7^2) + x_3(-m_3^2) + x_4(-m_4^2),
\end{eqnarray*}
and those (Eqs. (\ref{eq:D-cross}), (\ref{eq:C-cross})) for non-planar diagram (Fig.~\ref{fig:tlb-b-appen}) from
\begin{eqnarray*}
&&{{\mathsf A}}_{11}=x_1+x_2+x_3+x_4+x_5, \qquad{{\mathsf A}}_{12}={{\mathsf A}}_{21}=x_1+x_2+x_3, \\
&&{{\mathsf A}}_{22}=x_1+x_2+x_3+x_6+x_7,\\
&&{{\mathsf B}}_1=-(x_1+x_4)p_3-x_3(p_1+p_3)+x_2p_4, \qquad{{\mathsf B}}_2=(x_2+x_6)p_4-x_3(p_1+p_3)-x_1p_3,\\
&&{\mathsf c}=x_1(p_3^2-m_1^2)+x_4(p_3^2-m_4^2)+x_2(p_4^2-m_3^2)+x_6(p_4^2-m_6^2) \\
&&\qquad\qquad+ x_3((p_1+p_3)^2-m_3^2) + x_5(-m_5^2) + x_7(-m_7^2).
\end{eqnarray*}

\bibliographystyle{elsarticle-num}
%%%%%%%%% Bibliography  %%%%%%%%%%

%-----------------------------------------------------------------------

\end{document}